\documentclass[12pt]{article}
\usepackage{graphicx}
\def\ffrac#1#2{\textstyle{#1\over#2}\displaystyle}

\begin{document}
\parindent 0mm
\parskip 6pt
\title{SLE$(\kappa,\vec\rho\,)$ and Conformal Field Theory}
\author{John Cardy\\
Rudolf Peierls Centre for Theoretical Physics\\
         1 Keble Road, Oxford OX1 3NP, U.K.}
\date{January 13, 2006}
\maketitle
\begin{abstract}
SLE$(\kappa,\vec\rho\,)$ is a generalisation of Schramm-Loewner evolution which 
describes planar curves which are 
statistically self-similar but not conformally invariant in the strict sense.
We show that, in the context of boundary conformal field theory, this process
arises naturally in models which contain a conserved U$(1)$ current density
$J^\mu$, in which case it gives rise to a highest weight state $|h\rangle$
satisfying a deformation
$2L_{-2}|h\rangle=(\kappa/2)L_{-1}^2|h\rangle+\alpha J_{-1}L_{-1}|h\rangle$
of the usual level 2 null state condition. 

We apply this to a free field theory with piecewise constant Dirichlet
boundary conditions, with a discontinuity $\lambda$ at the origin, and
argue that this will lead to level
lines in the bulk described by SLE$(4,\vec\rho\,)$ across which 
there is a universal
macroscopic jump $\pm\lambda^*$ in the field, independent of the value
of $\lambda$.
\end{abstract}
\newpage
\section{Introduction}
\label{sec1}
In recent years, Schramm-Loewner evolution (SLE)\cite{Sc,LSW,RS,REV} has 
revolutionised the study of the continuum limit of two-dimensional critical
systems. This approach focuses on finding the correct probability measure
to describe the random curves, such as cluster boundaries, which occur in
such systems. In the simplest setting of curves which connect two
distinct points $z_1$, $z_2$
on the boundary of a simply connected domain $\cal D$, this measure
is generated dynamically by evolving the curve starting from one end point.
Conventionally the domain is taken to be the upper half
plane $\bf H$, and the curve $\gamma_t$
as evolved up to time $t$ (or rather its hull
$K_t$, which includes any regions enclosed by the curve) is characterised
by the conformal mapping $g_t: {\bf H}\setminus K_t\to{\bf H}$, which is
unique if we demand that $g_t(z)\sim z+2t/z+O(z^{-2})$ at infinity. This
function satisfies the Loewner equation
$$
{dg_t(z)\over dt}={2\over g_t(z)-W_t}\,,
$$
where the continuous function
$W_t$ is the image of the growing tip of the curve under $g_t$.
In this paper we shall find it more useful to define
$\hat g_t(z)\equiv g_t(z)-W_t$, which always maps the growing tip of the
curve to the origin, and satisfies $d\hat g_t=2dt/\hat g_t-dW_t$.

This is completely general. However, Schramm\cite{Sc}
argued that if the following
conditions hold:
\begin{itemize}
\item{A.} If $\cal D$ is mapped conformally to another domain ${\cal D}'$,
so that a curve $\gamma$ is mapped to $\gamma'$, then the induced measure
on $\gamma'$ is the correct measure for $\gamma'$ in 
${\cal D}'$;
\item{B.} If $\gamma_t$ is the part of the curve up time $t$, then the
conditional measure on the remainder $\gamma\setminus\gamma_t$ in 
$\cal D$ is the same as the unconditional measure in ${\cal D}\setminus K_t$;
\end{itemize}
together with a reflection property, then the only possibility for the
driving term $W_t$ is Brownian motion, namely $dW_t=\sqrt\kappa dB_t$,
where $B_t$ is standard Brownian motion and $\kappa$ is the diffusion
constant. Different values of $\kappa$ correspond to different universality
classes of critical behaviour. 

The continuum limit of isotropic short-range 2d critical systems is also
believed to be described by conformal field theory (CFT). This focuses
on the correlation functions of local operators which are the scaling limits
of local lattice observables. Within radial quantisation and the
operator-state correspondence of CFT, these correspond to states in the
Hilbert space of the theory which fall into highest weight representations
of the Virasoro algebra. This approach appears to be very different from
that of SLE. Nevertheless it can be argued (see later)
that to each partial curve $\gamma_t$
can be associated a state in this space. The ensemble of curves satisfying
conditions (A) and (B) then corresponds to a highest weight state $|h\rangle$
whose Verma module contains a level 2 null state:
$\big(L_{-2}-(\kappa/4)L_{-1}^2\big)|h\rangle=0$.
This implies, using the BPZ\cite{BPZ} equations, 
that expectation values of all observables 
which can be written as $\langle{\cal O}|h\rangle$ satisfy
linear second-order differential equations. These are the same equations
found from the stochastic approach. This partial unification between the
two approaches has been very fruitful, both in understanding CFT
and in suggesting possible generalisations of SLE to 
other domains\cite{F} and to multiple curves\cite{JCmult}.

One of the important properties of Brownian motion is scale invariance:
$\sigma^{-1}W_{\sigma^2t}$ has the same law as $W_t$.
This 
extends to the whole sequence of Loewner mappings: 
$\sigma^{-1}g_{\sigma^2t}(\sigma z)$
obeys the same equation as $g_t$. This means that the curves described
by this process are statistically self-similar (which of course is a special
case of conformal symmetry.) SLE$(\kappa,\vec\rho\,)$ is a minimal way of
generalising SLE while retaining 
self-similarity. This is done by extending the stochastic equation
satisfied by $W_t$ to the system
\begin{eqnarray}
dW_t&=&\sqrt\kappa dB_t-\sum_{j=1}^n{\rho_jdt\over X^{(j)}_t}\,;\label{kr1}\\
dX^{(j)}_t&=&{2dt\over X^{(j)}_t}-dW_t\,,\label{kr2}
\end{eqnarray}
where the parameters $\vec\rho\,\equiv(\rho_1,\ldots,\rho_n)$ are constants.
The auxiliary variables $X^{(j)}_t=\hat g_t(X^{(j)}_0)$ 
are not independent of $W_t$: indeed
they could be integrated out. 
The curve $\gamma$ will thus depend not only on 
the particular realisation of the
Brownian motion but also the initial values $Z^{(j)}_0$ as well as 
the $\{\rho_j\}$.

SLE$(\kappa,\vec\rho\,)$ processes were first introduced in Ref.~\cite{D14}
as examples of restriction measures. Their properties have been
studied further in \cite{D7,D25}, and they were used in \cite{D}
in order to study
certain properties of SLE which also lead to a proof of 
Watts' formula\cite{Watts} for percolation.

Let us first remark that the equations (\ref{kr1},\ref{kr2}) are just special
cases of a more compactly stated problem. Let $\Phi_t(z)$ be the (unique up
to an additive constant)
harmonic function in the upper half plane which, on the real axis, is
piecewise constant with discontinuities $\pi\rho_j$ at the points
$X^{(j)}_t$. 
Define the current density ${\cal J}_t^\mu\equiv
\epsilon^{\mu\nu}\partial_\nu\Phi_t$,
which is conserved everywhere except at the sources $X^{(j)}_t$.
Then (\ref{kr1}) may be written
\begin{equation}
\label{kr3}
dW_t=\sqrt\kappa dB_t-{\cal J}_t^x(0)dt\,.
\end{equation}
Note that while $\Phi_t$ is conformally invariant, in the sense that
$\Phi_t({\hat g}^{-1}_t(z))$ is harmonic in ${\bf H}\setminus K_t$,
the measure on curves given by
(\ref{kr3}) is \em not \em invariant,
because the derivative in ${\cal J}^x_t=\partial_y\Phi_t$ is
with respect to $z$, not ${\hat g}^{-1}_t(z)$. 

We shall use (\ref{kr3}) as a general definition of 
SLE$(\kappa,\rho)$.\footnote{In this formulation the parameters $\vec\rho\,$
are hidden. However in this paper the original name will be used even
when the sources are not specified.} 
One advantage is that it may be used to define 
this process in other domains $\cal D$, where $\hat g_t$ maps 
${\cal D}\setminus K_t$ onto $\cal D$, and the current density ${\cal J}^\mu$ is
evaluated in this domain. Moreover, as will become apparent,
there are situations when (\ref{main}) applies even though
there are no sources for ${\cal J}^\mu$.

In this paper we extend the correspondence between random curves and
CFT to include those described by SLE$(\kappa,\rho)$, as defined by
(\ref{kr3}). It turns out that
the null state condition is replaced by a more general one
\begin{equation}
\label{main}
\big(2L_{-2}-(\kappa/2)L_{-1}^2-{\cal J}_{-1}L_{-1}\big)|h\rangle=0\,,
\end{equation}
where ${\cal J}_{-1}={\cal J}^x_0(0)$.
This is described in Sec.~\ref{sec2}.

These equations are more likely to be physically relevant
if the current density ${\cal J}^\mu$ arises naturally
in the theory under consideration. A simple example is the theory of
a free gaussian field $\phi$, with action $S=(g/4\pi)\int(\partial\phi)^2d^2z$
and with piecewise constant Dirichlet
boundary conditions. This is described in Sec.~\ref{sec3}.
The jumps in the boundary condition act as sources for the current
density
$J^\mu\propto\epsilon^{\mu\nu}\partial_\nu\phi$.
In particular we consider the effect of introducing
a jump of strength $\lambda$ at the origin, corresponding to the insertion of 
a boundary condition changing operator $\phi_\lambda(0)$. We show that
there is a special value $\lambda^*=(4g)^{-1/2}$ 
for which the corresponding highest
weight state satisfies the usual level 2 null state condition with 
$\kappa=4$. Its conformal weight $h=g{\lambda^*}^2$ takes the value $\frac14$.
For other values of $\lambda$, however, it satisfies 
the deformed condition (\ref{main}), with ${\cal J}=\alpha J$ where
$\alpha=\big((\lambda^*/\lambda)-(\lambda/\lambda^*)\big)$,
and $J$ normalised so that a discontinuity $\lambda^*$ corresponds
to unit charge. 

Of course, in this theory, changing $\lambda$ is equivalent to changing 
the coupling constant $g$, which is the same as perturbing the action
with the exactly marginal operator $(\partial\phi)^2
\propto J\bar J$. From this point of view, the perturbation partially screens
the current-current correlations. We show that this screening acts to 
renormalise the effective U$(1)$ charge $\lambda$ to the critical value 
$\lambda^*$, with effective conformal weight $\frac14$.

In Sec.~\ref{sec32} we consider the physical implications of this. The free
field theory should correspond to the continuum limit of a gaussian
free field on a lattice, for which it is possible to identify uniquely
the level lines of the field. For this theory
we argue that the deformed condition (\ref{main})
is satisfied by the state corresponding to curves which are the
level lines of the free field emanating from the discontinuity $\lambda$
in $\phi$ at the origin. If $\lambda=\pm\lambda^*$ it does not matter which
level line, because there is a macroscopic jump $\pm\lambda^*$ in the field
$\phi$ across the curve in the bulk. 
This curve should be described by SLE$_4$, as proved by Sheffield and
Schramm\cite{SS}. 

We shall argue that if $\lambda\not=\lambda^*$ then the level lines
are described by SLE$(\kappa,\vec\rho\,)$. However, the screening alluded
to above has the effect of renormalising the effective jump across the
curve in the bulk to the universal value $\lambda^*$, rather than
$\lambda$. The physics of this is clearer in
the discrete gaussian model, where the
values of $\phi(r)$ at lattice sites are integer multiples of some unit
$2\pi\Lambda$. 

\section{Random curves and CFT}
\label{sec2}

We start by discussing how the measure on curves $\gamma$ can be used 
generate a highest weight 
state in boundary CFT. Our point of view is slightly
different from that of Ref.~\cite{BB}, being better suited to the generalisation
to SLE$(\kappa,\vec\rho\,)$. In BCFT, we suppose that 
there is some set of fundamental local fields $\psi(r)$ (the continuum
limit of the local lattice degrees of freedom), satisfying given
conformally invariant boundary conditions on the real axis (with the possible
exception of the origin) and with a unnormalised
Gibbs measure $e^{-S[\psi]}[d\psi]$. The Hilbert space is that
of all possible field configurations $\psi_\Gamma$ on a fixed semicircle
$\Gamma$ centred on the origin.  The vacuum state
is given by weighting each state $|\psi'_\Gamma\rangle$
by the (normalised) path integral restricted to the interior of $\Gamma$
and conditioned on the fields taking 
the specified values $\psi'_\Gamma$ on the boundary:
$$
|0\rangle=\int[d\psi'_\Gamma]\int_{\psi_\Gamma=\psi'_\Gamma}
[d\psi]\,e^{-S[\psi]}\,|\psi'_\Gamma\rangle\,.
$$
Similarly, inserting a local operator $\phi(0)$ at the origin into
the path integral defines a state $|\phi\rangle$. 
This is the well-known
operator-state correspondence of CFT. Because of scale invariance,
a different choice for $\Gamma$ gives the same states, up to a multiplicative
constant.

Now suppose that for every field
configuration $\psi$ we can
identify a curve $\gamma$ connecting the origin to infinity. The existence
of such a curve is assumed to be guaranteed by the boundary conditions
on the real axis. On the lattice,
for example if $\gamma$ is a cluster boundary in the Ising model, we take
the spins on the negative real axis to be $-1$, and those on the
positive real axis to be $+1$. We assume this property continues to hold
in the continuum limit.
Any such curve may be generated by a Loewner process: denote as before
the part of the curve up to time $t$ by $\gamma_t$. 
The existence of this curve depends on only the field 
configurations $\psi$ in the interior of $\Gamma$, as long as $\gamma_t$
lies wholly inside this region.
Then we can condition the fields contributing to the path integral 
on the existence of $\gamma_t$, thus defining a state
$$
|\gamma_t\rangle =
\int[d\psi'_\Gamma]\int_{\psi_\Gamma=\psi'_\Gamma;\gamma_t}
[d\psi]\,e^{-S[\psi]}\,|\psi'_\Gamma\rangle\,.
$$
The path integral (over the whole of the upper half plane, not just the 
interior of $\Gamma$), when conditioned on $\gamma_t$, gives a measure
$d\mu(\gamma_t)$. The state
$$
|h\rangle=|h_t\rangle\equiv\int d\mu(\gamma_t)|\gamma_t\rangle
$$
is in fact independent of $t$, since it is just given by the path
integral conditioned on there being a curve connecting the origin to
infinity, which is
guaranteed by the boundary conditions. In fact, by taking $t=0$, we
see that $|h\rangle$ is just the state corresponding to
a boundary condition changing operator\cite{JCbcc} at the origin.

However, $d\mu(\gamma_t)$ is also given by the measure on
$W_t$ in Loewner evolution, through the iterated sequence of conformal
mappings satisfying $d\hat g_t=2dt/\hat g_t-dW_t$. 
This corresponds
to an infinitesimal conformal mapping of the upper half plane
minus $\gamma_t$.
In CFT, this is implemented by considering a more
general transformation $x^\mu\to x^\mu+\alpha^\mu(x)$, where
$\alpha^\mu(x)$ agrees with the conformal transformation inside a contour
$C$ which surrounds $K_t$ but lies inside $\Gamma$,
but vanishes outside $C$. This is compensated by inserting 
$(1/2\pi i)\int_C\alpha(z)T(z)dz+{\rm c.c.}$ into the path integral, 
where $T(z)$ is the local stress tensor.
Because of
the conformal boundary condition that $T=\overline T$ on the real axis,
we may drop the second c.c. term by extending $C$ to include its
reflection in the real axis.
In our case, $d\hat g_t$ corresponds to inserting
$(1/2\pi i)\int_C(2dt/z-dW_t)T(z)dz$.
In operator language, this corresponds to acting on $|\gamma_t\rangle$
with $2L_{-2}dt-L_{-1}dW_t$ where $L_n=(1/2\pi i)\int_Cz^{n+1}T(z)dz$. 
Thus, for any $t_1<t$,
$$
|g_{t_1}(\gamma_t)\rangle
={\bf T}\exp\left(\int_0^{t_1}\big(2L_{-2}dt'-L_{-1}dW_{t'}\big)\right)
|\gamma_t\rangle\,,
$$
where $\bf T$ denotes a time-ordered exponential. 

The measure on $\gamma_t$ is the product of the
measure of $\gamma_t\setminus\gamma_{t_1}$, conditioned on
$\gamma_{t_1}$, with the unconditioned measure on $\gamma_{t_1}$. 
The first is the same
as the unconditioned measure on $g_{t_1}(\gamma_t)$, and the second is given
by the measure on $W_{t'}$ for $t'\in[0,t_1]$. 
Thus
$$
|h_t\rangle=\int d\mu(g_{t_1}(\gamma_t))\int d\mu(W_{t';t'\in[0,t_1]})
{\bf T}e^{-\int_0^{t_1}\big(2L_{-2}dt'-L_{-1}dW_{t'}\big)}
|g_{t_1}(\gamma_t)\rangle\,.
$$
For ordinary SLE, $W_t$ is proportional to a Brownian process. The
integration over realisations of this for $t'\in[0,t_1]$ may be performed
by breaking up the time interval into small segments, expanding out
the exponential, using $(dB_{t'})^2=dt'$, and re-exponentiating. 
The result is
$$
|h_t\rangle=\exp\left(-\big(2L_{-2}-(\kappa/2)L_{-1}^2\big)t_1\right)
|h_{t-t_1}\rangle\,.
$$
But, as we argued from the path integral, $|h_t\rangle$ is
independent of $t$, and therefore
\begin{equation}
\label{l2l1}
\big(2L_{-2}-(\kappa/2)L_{-1}^2\big)|h\rangle=0\,,
\end{equation}
that is, there is a level 2 null state. 
Note that $|h\rangle=|h_0\rangle$ is also expected to be a
highest weight state, $L_n|h\rangle=0$ ($n\geq1$), since in this case
we can shrink $C$ to zero.

For SLE$(\kappa,\rho)$ the measure depends on the values of 
$\{X_0^{(j)}\}$ and therefore so does the state
$|h_t;\{X_0^{(j)}\}\rangle$. In writing how this state behaves under
the infinitesimal conformal mapping $d\hat g_t$, we need to decide
whether the points $\{X_t^{(j)}\}$ lie inside or outside the contour $C$.
In order to be able to define a highest weight state, we need to be able
to shrink $C$ to the origin without obstructions, so it is natural to choose
it to lie inside all the $\{X_t^{(j)}\}$. But this has the (convenient)
consequence that these points do not evolve under the modified transformation
which vanishes outside $C$. Thus the drift term in (\ref{kr1}) is constant,
and it is still straightforward to integrate over the measure on $dB_{t'}$,
to obtain

$$
|h_t;\{X^{(j)}_0\}\rangle=
\exp\Big(-\big(2L_{-2}-(\kappa/2)L_{-1}^2+
\sum_j(\rho_j/X^{(j)}_0)L_{-1}\big)t_1\Big)
|h_{t-t_1};\{X^{(j)}_0\}\rangle\,,
$$
so we can argue, as before, that there is a time-independent state
$|h;\{X^{(j)}_0\}\rangle$ satisfying
\begin{equation}
\label{hX}
\Big(2L_{-2}-(\kappa/2)L_{-1}^2+
\sum_j(\rho_j/X^{(j)}_0)L_{-1}\Big)|h;\{X^{(j)}_0\}\rangle=0\,.
\end{equation}

Note that if we had taken $C$ to lie outside all the $X^{(j)}_t$
this would incur the replacement
\begin{eqnarray*}
L_{-2}&\rightarrow& L_{-2}+\sum_j(2/X^{(j)}_{t'})\partial_{X^{(j)}_{t'}}\,,\\
L_{-1}&\rightarrow& L_{-1}+\sum_j\partial_{X^{(j)}_{t'}}\,,
\end{eqnarray*}
corresponding to diffusion in the moduli space of the half-plane
with marked boundary points, as well as the usual diffusion\cite{F}.
In addition,
the evolved state would not be of highest weight.
There is no contradiction here: the points $\{X^{(j)}_{t}\}$
act as the location of boundary condition changing operators 
$\Phi_j(X^{(j)}_{t})$, and the state we get by taking $C$ to lie outside
these points is not of highest weight and transforms non-trivially.

The last term in (\ref{hX}) may be written in terms of the current
density
${\cal J}(z)=\sum_j\rho_j/(z-X^{(j)}_0)$ introduced in Sec.~\ref{sec1}.
Its value at the origin is ${\cal J}_{-1}=(1/2\pi i)\int_Cz^{-1}J(z)dz$.
This gives (\ref{main}).

\section{Free field theory}
\label{sec3}

In this section we illustrate the above for the simplest case
of a free field theory.
Consider field $\phi(r)$ in the upper half plane with action
$S[\phi]=(g/4\pi)\int(\partial\phi)^2d^2z$.
The boundary conditions are piecewise Dirichlet: however there are
discontinuities with jumps $2\pi\lambda_j$ at points $x_j$ with
$j=(0,1,\ldots,n)$:
$$
\phi(x)=2\pi\sum_j\lambda_j H(x-x_j)\,,
$$
where $H(x)=1$ for $x>0$ and 0 for $x<0$. 
These boundary conditions are satisfied by the harmonic function
$$
\phi_c(z,\bar z)=-2\sum_j\lambda_j\,{\rm arg}(z-x_j)
=i\sum_j\lambda_j\ln\big((z-x_j)/(\bar z-x_j)\big)\,,
$$
and if we write $\phi=\phi_c+\phi'$, with $\phi'=0$ on the boundary,
$S[\phi]=S[\phi_c]+S[\phi']$, so that the partition function is
$Z=Z_\lambda\,Z'$, where 
\begin{equation}
\label{Zc}
Z_\lambda=\prod_{j<k}\big((x_k-x_j)/a\big)^{2g\lambda_j\lambda_k}\,,
\end{equation}
where $a$ is the UV cutoff, and $Z'$ is the partition function for
homogeneous Dirichlet boundary conditions. 

In BCFT\cite{JCbcc}, we can think of $Z_\lambda/Z'$ as the correlation function
$\langle\prod_j\phi_{\lambda_j}(x_j)\rangle$ of boundary condition changing
operators. Thus any expectation value $\langle{\cal O}\rangle$
of some observable with the inhomogenous boundary conditions can be
written
$$
\langle{\cal O}\rangle_\lambda={\langle{\cal O}\prod_j\phi_{\lambda_j}
(x_j)\rangle \over \langle\prod_j\phi_{\lambda_j}(x_j)\rangle}\,.
$$

In particular, we can compute the expectation value of the stress tensor
$$
\langle T(z)\rangle_\lambda=-g(\partial_z\phi_c)^2
=g\sum_j\sum_k{\lambda_j\lambda_k\over(z-x_j)(z-x_k)}\,,
$$
and compare its behaviour as $z\to x_i$ with that expected from the
conformal Ward identity\cite{BPZ}
$$
T(z)\,\phi_{\lambda_i}(x_i)={h_i\over(z-x_i)^2}\phi_{\lambda_i}(x_i)
+{1\over z-x_i}\partial_{x_i}\phi_{\lambda_i}(x_i)+
L_{-2}\phi_{\lambda_i}(x_i)+O(z-x_i)\,.
$$
An explicit computation gives
\begin{eqnarray*}
\langle T\rangle_\lambda&=&
{g\lambda_i^2\over(z-x_i)^2}+{2g\over z-x_i}
\sum_j{}'{\lambda_i\lambda_j\over x_i-x_j}\\
&&\qquad\qquad -2g\sum_j{}'{\lambda_i\lambda_j\over(x_i-x_j)^2}
+g\sum_{j,k}{}'{\lambda_j\lambda_k\over(x_i-x_j)(x_i-x_k)}+
O(z-x_i)
\end{eqnarray*}
where a prime on the sum omits the terms with $j,k=i$.
From this we identify the scaling dimension of $\phi_{\lambda_i}$ to be
$$
h_i=g\lambda_i^2\,,
$$
and see that
$$
\partial_{x_i}\ln\langle\prod_j\phi_{\lambda_j}(x_j)\rangle
=2g\sum_j{}'{\lambda_i\lambda_j\over x_i-x_j}\,,
$$
which is of course consistent with (\ref{Zc}). 

We also see that
$$
\langle L_{-2}\phi_{\lambda_i}(x_i)\prod_j{}'\phi_{\lambda_j}(x_j)\rangle
=-2g\sum_j{}'{\lambda_i\lambda_j\over (x_i-x_j)^2}
+g\sum_{j,k}{}'{\lambda_j\lambda_k\over(x_i-x_j)(x_i-x_k)}\,,
$$
while
\begin{eqnarray*}
\langle L_{-1}^2\phi_{\lambda_i}(x_0)
\prod_j{}'\phi_{\lambda_j}(x_j)\rangle&=&
\prod_j{}'(x_i-x_j)^{-2g\lambda_i\lambda_j}\,
\partial^2_{x_i}
\prod_j{}'(x_i-x_j)^{2g\lambda_i\lambda_j}\\
&=&-2g\sum_j{}'{\lambda_i\lambda_j\over (x_i-x_j)^2}
+4g^2\sum_{j,k}{}'{\lambda_i^2\lambda_j\lambda_k\over(x_i-x_j)(x_i-x_k)}
\,.
\end{eqnarray*}

The condition $2L_{-2}\phi_{\lambda_i}=(\kappa/2)L_{-1}^2\phi_{\lambda_i}$
is satisfied (as an operator condition, that is for all choices
of the $\lambda_j$ and $x_j$), only if
$$
\kappa=4\qquad\mbox{and}\qquad \lambda_i^2={\lambda^*}^2=1/4g\,,
$$
so that $h_i=\frac14$.

If the latter condition is not satisfied
we have instead
\begin{equation}
\label{deform}
2L_{-2}\phi_{\lambda_i}=2L_{-1}^2\phi_{\lambda_i}-
\sum_j{}'{\rho_j\over x_j-x_i}\,L_{-1}\phi_{\lambda_i}\,,
\end{equation}
where 
$$
\rho_j=(\lambda_j/\lambda_i)\left(1-(\lambda_i/\lambda^*)^2\right)\,.
$$

This has the form of (\ref{hX}), realised on operators rather than states,
with $x_i=0$, $X^{(j)}_0=x_j-x_i$. 
It is equally unsatisfactory as a local condition on $\phi_{\lambda_i}$, 
but can be written, as (\ref{main}), in terms of ${\cal J}_{-1}$. However
in this theory, this is, up to a multiplicative factor, the
conserved U$(1)$ current density
$J^\mu\propto\epsilon^{\mu\nu}\partial_\nu\phi$, for which the 
discontinuities at the boundaries at as local sources. It is useful 
to normalise this current so that
$$
J(z)\phi_{\lambda^*}(0)\sim(1/z)\phi_{\lambda^*}(0)
$$
(where $J=J_z$), so that a jump of $\lambda^*$ has unit U$(1)$ charge.
This means taking  $J=-2ig^{1/2}\partial_z\phi$. 

As for $T(z)$, we can define the modes $J_n=(1/2\pi i)\int_Cz^nJ(z)dz$.
The term in $\partial_{x_i}\phi_c$ with $j=i$ corresponds to $J_0$:
the rest is $J_{-1}$. (\ref{deform}) may then be rewritten as
an operator condition on a highest weight state $|h_\lambda\rangle$:
\begin{equation}
\label{m2}
\big(2L_{-2}-2L_{-1}^2-\alpha J_{-1}L_{-1}\big)
|h_\lambda\rangle\,,
\end{equation}
where $\alpha=q_\lambda^{-1}-q_\lambda$ with
$q_\lambda=\lambda/\lambda^*$. 

Although (\ref{m2}) has been derived for the specific case when the current
$J$ is produced by sources which are themselves boundary condition changing
operators, it is valid completely generally as an operator condition. 
With the normalisation of the current chosen above 
we have $T=-g:\!(\partial_z\phi)^2\!:
=\frac14:\!J^2\!:$ where $J\equiv J_z$. Thus, in terms of operators,
$$
L_n=\ffrac14\sum_r:\!J_rJ_{n-r}\!:\,,
$$
where now the normal ordering symbol $:\!J_kJ_l\!:$
places $J_k$ to the right of $J_l$ if $k>l$. Let $|h,q\rangle$ be a highest
weight state of conformal weight $h$ and U$(1)$ charge $q$, so that
$L_0|h,q\rangle=h|h,q\rangle$, $J_0|h,q\rangle=q|h,q\rangle$, and
$L_n|h,q\rangle=J_n|h,q\rangle=0$ for $n>0$.
Then
\begin{eqnarray*}
L_{-2}|h,q\rangle&=&\ffrac14\big(2J_{-2}J_0+J_{-1}^2\big)|h,q\rangle=
\ffrac14\big(2qJ_{-2}+J_{-1}^2\big)|h,q\rangle\\
L_{-1}|h,q\rangle&=&\ffrac14\big(2J_{-1}J_0+2J_{-2}J_1\big)|h,q\rangle=
\ffrac14(2qJ_{-1})|h,q\rangle\\
L_{-1}^2|h,q\rangle&=&
\ffrac1{16}\big(2J_{-1}J_0+2J_{-2}J_1\big)(2qJ_{-1})|h,q\rangle=
\ffrac14\big(q^2J_{-1}^2+kqJ_{-2}\big)|h,q\rangle\\
J_{-1}L_{-1}|h,q\rangle&=&\ffrac12qJ_{-1}^2|h,q\rangle\,,
\end{eqnarray*}
where we have introduced the U$(1)$ anomaly $k$ by
$\langle J(z)J(0)\rangle=k/z^2$, so that $[J_n,J_m]=kn\delta_{n,-m}$.
An explicit calculation gives $k=2$ with this normalisation for $J$, so that
on forming the difference $(2L_{-2}-2L^2_{-1})|h,q\rangle$, we again obtain
(\ref{m2}), but completely generally.

Note that in the whole of this section, instead of considering Dirichlet
boundary conditions with boundary condition changing operators 
$\phi_{\lambda}$, we could, in the dual description, have considered
Neuman boundary conditions on the dual field $\tilde\phi$,
with insertions of vertex operators $e^{i\lambda\tilde\phi}$. 

\subsection{$J\bar J$ perturbation}
\label{sec31}
Let us make the simple observation that if
$\lambda\not=\lambda^*(g)=1/2g^{1/2}$, we can always make it so by suitably
changing $g$. Pick some reference value $g_0$, so that action
is $S_0=(g_0/4\pi)\int(\partial\phi)^2d^2z$. With respect to this action,
a jump of $\lambda^*=1/2g_0^{1/2}$ is a unit source for the current
density with components
$J=-2ig_0^{1/2}\partial_z\phi$, $\bar J=2ig_0^{1/2}\partial_{\bar z}\phi$.
Now consider the perturbed action
$$
S=S_0+u\int J\bar Jd^2z=(g_{\rm eff}(u)/4\pi)
\int(\partial\phi)^2d^2z\,,
$$
where $g_{\rm eff}(u)=g_0(1+4\pi u)$. 

In the perturbed theory, the scaling dimension of $\phi_{\lambda^*}$
is modified to $g_{\rm eff}{\lambda^*}^2=\frac14(1+4\pi u)$.
The jumps corresponding to $h=\frac14$ are now
$\pm\lambda^*(1+4\pi u)^{-1/2}$. 

Furthermore, the current-current correlations are partially screened. In the
unperturbed theory 
$$
\langle J(z)J(0)\rangle=k/z^2\quad\mbox{and}\quad
\langle \bar J(\bar z)\bar J(0)\rangle=k/{\bar z}^2\,,
$$
where $k=2$.
This implies, for example, that
$$
\langle J_y(x,y)J_y(0,0)\rangle=-{x^2-y^2\over(x^2+y^2)^2}\,,
$$
so that two currents alongside each other are more likely to be anti-parallel
rather than parallel. This is because there are many small closed current
loops. In the perturbed theory, keeping the normalisation of the current
the same, the U$(1)$ anomaly is reduced (if $u>0$) by a factor
$(1+4\pi u)^{-1}$. This is as if the effective current along
a given loop were reduced by $(1+4\pi u)^{-1/2}$. 
This means that the effective charge of $\phi_{\lambda^*}$ corresponds
to a scaling dimension $\frac14$.

In the perturbed theory we now have $T=\frac14(1+4\pi u):\!J^2\!:$, so that
\begin{eqnarray*}
L_{-2}|h_\lambda\rangle&=&\ffrac14(1+4\pi u)\big(2qJ_{-2}+J_{-1}^2\big)
|h_\lambda\rangle\\
L_{-1}^2|h_\lambda\rangle&=&\ffrac14(1+4\pi u)^2\big(q^2J_{-1}^2
+k(u)qJ_{-2}\big)|h_\lambda\rangle\\
J_{-1}L_{-1}|h_\lambda\rangle&=&\ffrac12q(1+4\pi u)J_{-1}^2\,,
\end{eqnarray*}
where $k(u)=2/(1+4\pi u)$. Then
$$
\big(2L_{-2}-2L_{-1}^2\big)|h_\lambda\rangle=
\big(q_{\lambda}^{-1}-(1+4\pi u)q_\lambda\big)|h_\lambda\rangle\,.
$$
Thus when $u=0$ the state $|h_{\lambda^*}\rangle$ satisfies
the ordinary level 2 null condition, corresponding to SLE$_4$,
while for $u\not=0$ it satisfies
the deformed relation (\ref{m2}), corresponding to SLE$(4,\rho)$,
and its conformal weight is modified accordingly.
On the other hand $|h_\lambda\rangle$
with $\lambda=\lambda^*(1+4\pi u)^{-1/2}$ satisfies the undeformed
condition, for all $u$. 

However, for $u\not=0$ there is partial screening
(or anti-screening) which happens in such a way that the effective current
is that which would emerge from a boundary condition changing operator
with the universal scaling dimension $\frac14$.

\subsection{Universal jump across level lines}
\label{sec32}
This part is speculative in nature. Many of the conjectures have in
fact been proved by Schramm and Sheffield\cite{SS}.
So far we have shown only that the equations (\ref{l2l1},\ref{m2})
hold in CFT for boundary condition changing operators in a free field
theory, and that they have the same form as satisfied by the highest
weight states for SLE$_4$ if $\lambda=\lambda^*$ and for 
SLE$(4,\vec\rho\,)$ otherwise. We have not yet identified the curves which
are described by these SLEs. A natural conjecture is that these are the
level lines of the free field $\phi$. For a free gaussian field theory on a 
triangular lattice, we can condition the field $\phi(r)$ on the existence
of a curve $\gamma$ which is a level line of height $\phi_0$ 
by demanding that $\phi(r)>\phi_0$ for sites $r$ immediately to the right
of the curve and $\phi(r)<\phi_0$ for sites immediately to its left. 
If we impose a jump in the boundary conditions at the origin, 
there will be such a level line connecting the
origin to infinity for each value of $\phi_0$ satisfying
$\phi(0-)<\phi_0<\phi(0+)$. On the lattice these curves will in 
general be different, according to how $\phi_0$ is chosen.

Let us first consider the case $\lambda=\lambda^*$. Which of these curves
is a suitable candidate for SLE$_4$? The answer to this appears to be
that they all are the same curve in the continuum limit. Recall 
condition (B) for SLE, that the conditional measure on $\gamma\setminus
\gamma_t$ in $\cal D$ should be the same as the unconditioned measured
in ${\cal D}\setminus\gamma_t$ (here we assume that $\gamma_t$ is simple, so
$\gamma_t=K_t$). For this to be the case, the values of the field
$\phi$ on either side of $\gamma_t$ should take the same values
$\phi(0-)$ and $\phi(0+)$ as on the negative and positive real axes.
This is not necessarily the case in the lattice model, however. The
values of the field either side are conditioned only to be respectively
less than, or greater than, $\phi_0$.
Given that we have argued that for $\lambda=\lambda^*$ the state satisfies
the same equation in the continuum limit 
as does an SLE$_4$, it is natural to conjecture that
this particular value gives rise, in the continuum limit,
to level lines across which the jump
in the field is $\lambda^*$ everywhere. 
That is, conditioning the values on either side to take values either side of
$\phi_0$, together with the existence of a jump $\lambda^*$ at the boundary,
has the result of enforcing a macroscopic jump $\lambda^*$ all the way along
$\gamma$, in the continuum limit. 
Schramm and Sheffield\cite{SS} have in fact proved that the level lines
of the free field, defined as above on the lattice, converge to SLE$_4$
in the continuum limit, as long as $\lambda$ takes a particular value.
That the jump in the field across this curve is also $\lambda^*$ has been
verified in simulations by S.~Sheffield\cite{SS}.

If $\lambda\not=\lambda^*$, the equation satisfied by the boundary state
does not correspond to simple SLE, but we have shown that it does correspond
to SLE$(4,\vec\rho\,)$.
However we argued in the previous section this is equivalent to taking
$\lambda=\lambda^*$, at the same time perturbing the action by a term
$\propto\int{\vec J}\cdot{\vec J}d^2z$, and that this results in partial
screening
of the current-current correlations. The physics of screening is more
transparent when the charges are discrete rather than continuous.
For this reason let us consider a discrete gaussian model on the lattice,
in which the field $\phi(r)$ at each lattice site is an integer multiple
of some unit $2\pi\Lambda$. This may be enforced by adding to the action a
term $\int\cos(\phi/\Lambda)d^2z$, which has bulk scaling dimension
$x=1/(2g\Lambda^2)$. This is irrelevant if $x>2$, that is $\Lambda<\lambda^*$,
in which case the continuum limit of the discrete gaussian model is given
by a free field theory. On the hand if $\Lambda>\lambda^*$ the perturbation
is relevant, and the theory is no longer critical. For $\Lambda=\lambda^*$
it is in fact marginally irrelevant. 

\begin{figure}[t]
\centering
\includegraphics[width=10cm]{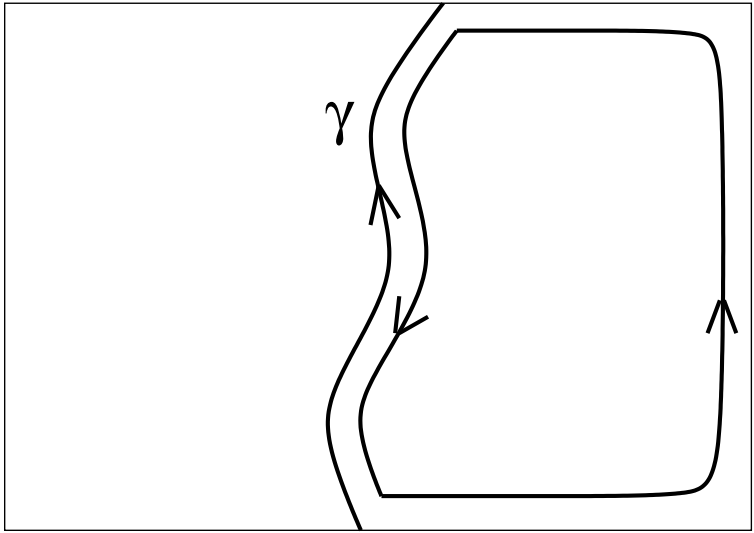}
\caption{\label{fig1}\small
An energetically favoured configuration if $u>0$, leading to 
screening of the
current along the curve $\gamma$.}
\end{figure}

Consider therefore a discrete gaussian model with $\Lambda=\lambda^*$.
On the lattice, the loops on the dual lattice carry integer
currents $\bf I$ (with the previous normalisation), 
and they intersect the boundary at points
corresponding to boundary operators with integer charges.
Such a curve with current $\pm1$ will correspond to a
state satisfying (\ref{l2l1}) and therefore should be described by SLE$_4$. 
Now switch on the current-current interaction. This may be modelled
on the lattice by a short-range interaction $u\sum_{R,R'}f(R-R')
{\bf I}(R)\cdot{\bf I}(R')$ between the currents on nearby loops. 
As argued in Sec.~\ref{sec2}, this will lead to partial
screening if $uf>0$: a current 
$\bf I$ will attract those parts of nearby loops with currents anti-parallel
to $\bf I$ and repel those parts with parallel currents. The resultant
effective current along the curve will be reduced by a 
factor $(1+4\pi u)^{-1/2}$. Similarly, there will be anti-screening if
$u<0$. Since the currents are discrete, this can of course
only happen in an average sense. In Fig.~\ref{fig1} we illustrate a loop
configuration which would contribute to the screening phenomenon. 
Note that in this case there are no sources for $\cal J$. If there were,
they could also contribute to the screening.
\em Acknowledgments\em. I would like to thank Scott Sheffield for explaining
to me some of the results of Ref.~\cite{SS} before publication, and 
Roland Friedrich for discussions. This work was carried out while the author
was a joint member of the Schools of Mathematics and Natural Sciences of
the Institute for Advanced Study, and was supported by the Ellentuck Fund.

\begin{thebibliography}{99}
%
\bibitem{Sc} O.~Schramm, Israel J. Math. {\bf 118}, 221, 2000.
%
\bibitem{LSW}
G.~Lawler, O.~Schramm and W.~Werner, Acta Mathematica {\bf 187} 237, 2001
(math.PR/9911084);
\em ibid. \em {\bf 187} 275, 2003 (math.PR/0003156);
Ann. Henri Poincar\'e {\bf 38} 109, 2002 (math.PR/0005294).
%
\bibitem{RS} S.~Rohde and O.~Schramm, Ann. Math., to appear (math.PR/0106036). 
%
\bibitem{REV} For reviews, see
W.~Werner, {\sl Random planar curves and Schramm-Loewner
evolutions}, to appear (Springer Lecture Notes)
(math.PR/0303354); G.~Lawler, {\sl Conformally
Invariant Processes in the Plane}, in preparation,
http://www.math.cornell.edu/\~lawler/book.ps;
W.~Kager and B.~Nienhuis, J. Stat. Phys. {\bf 115}, 1149 2004 
(math-ph/0312056);
J.~Cardy, {\sl SLE for theoretical physicists}, in preparation.
%
\bibitem{BPZ} A.A.~Belavin, A.M.~Polyakov and A.B.~Zamolodchikov,
Nucl. Phys. B {\bf 241}, 333, 1984.
%
\bibitem{F} R.~Friedrich, math-ph/0410029; R.~Bauer and R.~Friedrich,
math.PR/0408157.
%
\bibitem{JCmult} J.~Cardy, J. Phys. A {\bf 36}, L379, 2003
(erratum J. Phys. A {\bf 36}, 12343, 2003);
Phys. Lett. B {\bf 582}, 121, 2004.
%
\bibitem{D14} G.~Lawler, O.~Schramm and W.~Werner, J. Amer. Math. Soc.
{\bf 16(4)}, 917, 2003 (math.PR/0209343).
%
\bibitem{D7} J.~Dub\'edat, Ann. Probab., to appear (math.PR/0303128).
%
\bibitem{D25} W.~Werner, Ann. Fac. Sci. Toulouse, to appear
(math.PR/0302115).
%
\bibitem{D} J.~Dub\'edat, math.PR/0405074.
%
\bibitem{Watts} G.~Watts, J. Phys. A {\bf 29}, L363, 1996 (cond-mat/9603167).
%
\bibitem{SS} O.~Schramm and S.~Sheffield, in preparation; S.~Sheffield,
math.PR/0312099 and talk presented at `Conformal Invariance and Random 
Spatial Processes', Edinburgh, July 2003.
%
\bibitem{BB}  M.~Bauer and D.~Bernard, Comm. Math. Phys. {\bf 239}, 493,
2003 (hep-th/0210015); Phys. Lett. B {\bf 543}, 135, 2002;
Phys. Lett. B {\bf 557}, 309, 2003 (hep-th/0301064);
Ann. Henri Poincar\'e {\bf 5}, 289, 2004 (math-ph/0305061).
%
\bibitem{JCbcc} J.~Cardy, Nucl. Phys. B {\bf 324}, 581, 1989.
%
\bibitem{dNN} B.~Nienhuis, J. Stat. Phys. {\bf 34}, 731, 1983.
%
\bibitem{DF} Vl.~Dotsenko and V.~Fateev, Nucl. Phys. {\bf B240}, 312, 1984;
\em ibid. \em {\bf B251}, 691, 1985.
%
\bibitem{Kond} J.~Kondev, Phys. Rev. Lett. {\bf 78}, 4320, 1997.
%
\end{thebibliography}
\end{document}